# Chiral broadband High Harmonic Generation Source by Vectorial Time-Polarization-Gating

Eran Ben Arosh[1,2,†], Eldar Ragonis[1,2,†], Lev Merensky[1,2] and Avner Fleischer[1,2]*

[1] *Raymond and Beverly Sackler Faculty of Exact Sciences, School of Chemistry, Tel Aviv University, Tel Aviv 6997801, Israel*
[2] *Tel-Aviv University center for Light-Matter-Interaction, Tel Aviv 6997801, Israel*
† Equal contributors
*Corresponding Author: avnerfleisch@tauex.tau.ac.il*

**Abstract**
**Chiral (highly helical) extreme ultraviolet (XUV) sources are pivotal for investigating chiroptical phenomena on the ultrafast electronic timescale. Table-top, coherent High Harmonic Generation (HHG)-based sources are particularly well-suited for these studies. However, chiral materials, such as organic chiral molecules and solid-state magnetic materials, exhibit fine spectral features which necessitate broadband radiation for their complete interrogation. The generation of radiation that is both broadband and helical through HHG presents a seemingly paradoxical challenge: while chiral HHG emission requires at least two recollisions occurring along different directions in the polarization plane, the Floquet limit might already be reached with as few as three recollisions, resulting in a sparse spectrum characterized by pronounced discrete harmonic peaks. Notably, employing the well-established bicircular, two-color (ω, 2ω) driver scheme yields a sparse spectrum even with driver pulses close to single-cycle in duration, which restricts the detection of spectral features that vary on an energy scale finer than 1.55 eV for an 800 nm fundamental wavelength. Here we propose a straightforward scheme that enables the interrogation of fine spectral features, in principle restricted only by the resolution of the XUV spectrometer, with chiral XUV light. Our method is based on using a vectorial two-color driver with close central-frequencies with slight symmetry breaking. It integrates the time-gating and polarization-gating techniques to generate a vectorial driver which induces well-controlled bursts of recollisions, occurring along different directions in the polarization plane. The method satisfies the dual requirements of an XUV source which is both broadband and helical. We perform polarization scan and demonstrate that the broadband XUV radiation exhibits rapid modulations in its spectral ellipticity, and fast alternation in its spectral helicities. The phase of modulations could be controlled by introducing a slight symmetry breaking. This allows us to control and modulate the XUV polarization state, which should enable the detection of chiroptical signals with enhanced sensitivity. Our experimental results are fully supported by theoretical and numerical analysis, revealing the full anatomy of the scheme. We anticipate that our scheme will facilitate studies of high-resolution chiroptical phenomena is solids and chiral molecules, such as of vibronic circular dichroism, with the advantage of superior sensitivity and specificity.**

Chiral (namely circularly-polarized and highly elliptically-polarized) extreme-ultra-violet (XUV) and soft X-ray light is used in many fields of science due to its ability to investigate electronic and magnetic structure of chiral materials, with ultrafast resolution. Bright XUV radiation with adjustable polarization is routinely produced at free-electron laser (FEL) facilities [Suzuki2014,Lutman2016] by e.g., reflective phase retarders which convert linearly-polarized XUV light to circularly-polarized one. The bandwidth of the XUV light at those facilities is however narrow, as the X-ray pulses are emitted with pulse durations of few tens of femtoseconds. The large cost of such large-scale facilities limit their accessibility, which posses another problem. A table-top alternative to FEL facilities are setups utilizing High Harmonic Generation (HHG).

High Harmonic Generation has become a well-established technique in the last 30years, with widespread applications. It is based on the recollision of a continuum electron with the parent ion from which it was born by strong laser field ionization. In the recollision, an attosecond burst of broadband XUV radiation is emitted, whose polarization state is usually linear or elliptical with low ellipticity. Repeating the recollision process for even a small number (two or three) of times, (as is in the case when a driver pulse whose duration is larger than a single-cycle is used), results in the broad emission spectrum becoming structured and sparse, with pronounced spectral density around odd-integer multiples of the driver frequency.

Many applications of HHG however, such as time-resolved studies of ultrafast electron dynamics in magnetic materials [Zayko2021] benefit from highly-elliptical (highly-helical) chiral XUV light [Fan2015, Kfir2017, Willems2020, Heinrich2021,Azoury2019]. Indeed it has been demonstrated that synchronization of the recollision events such that they would occur along different directions in a plane, could yield any arbitrary polarization state for the XUV radiation and ellipticities of almost unity have been achieved [Fleischer2014, Huang2018, Bengs2021]. Most current sources of helical XUV spectrum are however sparse (in the form of a train of helical attosecond pulses) [Fleischer2014, Kfir2015, Hickstein2015, Bengs2021, Han2023]. Additionally, chiral gas-phase molecules could be studied by using them as the HHG generating media itself [Cireasa2015, Baykusheva2018]. However, chiral molecules or thin layers of magnetic solids have fine spectral features (especially near the ionization threshold [Beaulieu2018, Ferre2015]), e.g. those related to vibronic transitions, which vary on the energy scale of few tens of mili-eV, which the current sources of sparse helical XUV spectrum might not be able to resolve (see e.g., Fig 4 in [Baykusheva2019], where the ionization threshold falls on the fractional harmonic order 13.2). Hence, the study of chiral molecules would benefit from XUV sources which are not only helical, but also broadband, in order to be able to uncover closely-spaced spectral transitions. These two requirements are seemingly self-contradictory since for a broad XUV emission, a low number of recollisions is required but for maintaining high ellipticity the angle between subsequent recollision directions should be large. These two requirements are hard to achieve within a short driver pulse [Huang2018, Bengs2021]. For instance, the central technique nowadays to produce helical XUV radiation, uses a colinear, two-color, bicircular, counter-rotating HHG scheme employing a 800nm driver field and its second harmonic. In that scheme a driver as short as even a single-cycle (duration 2.67fs) already induces 3 recollisions, which bring the dynamics close to the Floquet limit [Fleischer2005, Lucchini2022]. As a result, the resulting highly-helical emission in still concentrated around integer harmonics, with spin angular momentum selection rules excluding harmonics which are multiples of three, with typical bandwidth of 1eV around each integer peak. In noncollinear schemes [Garcia2016, Huang2018], the appearance of a supercontinuum indicates the existence of a single strong recollision (with possible weaker, satellite recollisions) which yields low ellipticity values for the XUV radiation once the beam is relay-imaged and used in a downstream experiment because then the inhomogeneous spatial composition of the focus is revealed. This calls for a different HHG scheme, which could reconcile these two seemingly contradicting requirements.

Here we show how this can be achieved (see Fig.1). Short driver pulses, as generated e.g. by post-compression techniques [Khazanov2022], is not the only way by which the number of induced recollisions could be reduced. Time-gating (TG) [Fleischer2006, Merdji2007], in which a driver pulse with two close central frequencies creates a beat note which modulates the pulse envelope, effectively limiting the number of ionization (and recollision) events, is an alternative method. Due to the high nonlinearity of the HHG process, which is sensitive to the instantaneous amplitude of the electric field, this amplitude modulation effectively splits the original pulse into a serie of sub-pulses, or gates. In each gate a burst of several recollisions occurs while between gates no recollision takes place. This total reduction in the number of recollisions results in an increased emission bandwidth around each (previously-odd) harmonic peak. All recollisions within a single gate are preferentially aligned along the same direction, and the obtained XUV radiation is close to being linearly-polarized. For helical

XUV radiation at least two gates, directed along different directions, are necessary. We show that this can be achieved by taking the two colors to be cross-linearly-polarized, namely by making the two-color TG field vectorial (VTG) instead of scalar. With short-enough pulses, this results in two perpendicular gates. Since recollisions occurring along perpendicular directions emit XUV bursts which interfere with each other through the polarization state of the XUV radiation only, but not through the HGS, this method yields highly-helical XUV light, but without the accompanying spectral modulations. Hence, the XUV emission remains broadband, with structure in the HGS resulting from interference between recollisions within a single gate only. Of course, in case where too many gates (recollisions) take place, the Floquet limit might be reached, not only modulating the HGS but in fact inversely affecting the helicity of the XUV radiation, making alternating harmonics cross-linearly-polarized [Lerner2025]. To avoid this, we slightly break the symmetry of the cross-linear configuration, by introducing slight ellipticities to both colors [Bordo2020]. Through experimental polarization scan performed on the XUV radiation and numerical analysis, we show that this scheme generates helical XUV emission, keeping the spectral modulation to minimal. Importantly, we obtain periodic modulations of the field's helicity (from left-handed to right-handed) between nearby frequencies. In order to gain insight into the process, we continuously control the polarization state of the two colors and show how the onset of features dictated by selection rules imposed by spin angular momentum considerations [Fleischer2014] could be observed. This persistent adhering of the scheme to the selection rules imposed by the Floquet limit, in spite of the fact of the driver field is far from being strictly time-periodic, makes our scheme especially relevant for the detection of ultrafast dynamics in chiral matter. This is because the periodic pattern of helicity modulation constitute a reference against which chiral measurements could be performed, in the same way that lock-in measurements are done using acousto-optic modulators in circular dichroism (CD) measurements in the Visible regime. This should increase the sensitivity of our XUV source in the detection of chiral matter as compared to other methods.

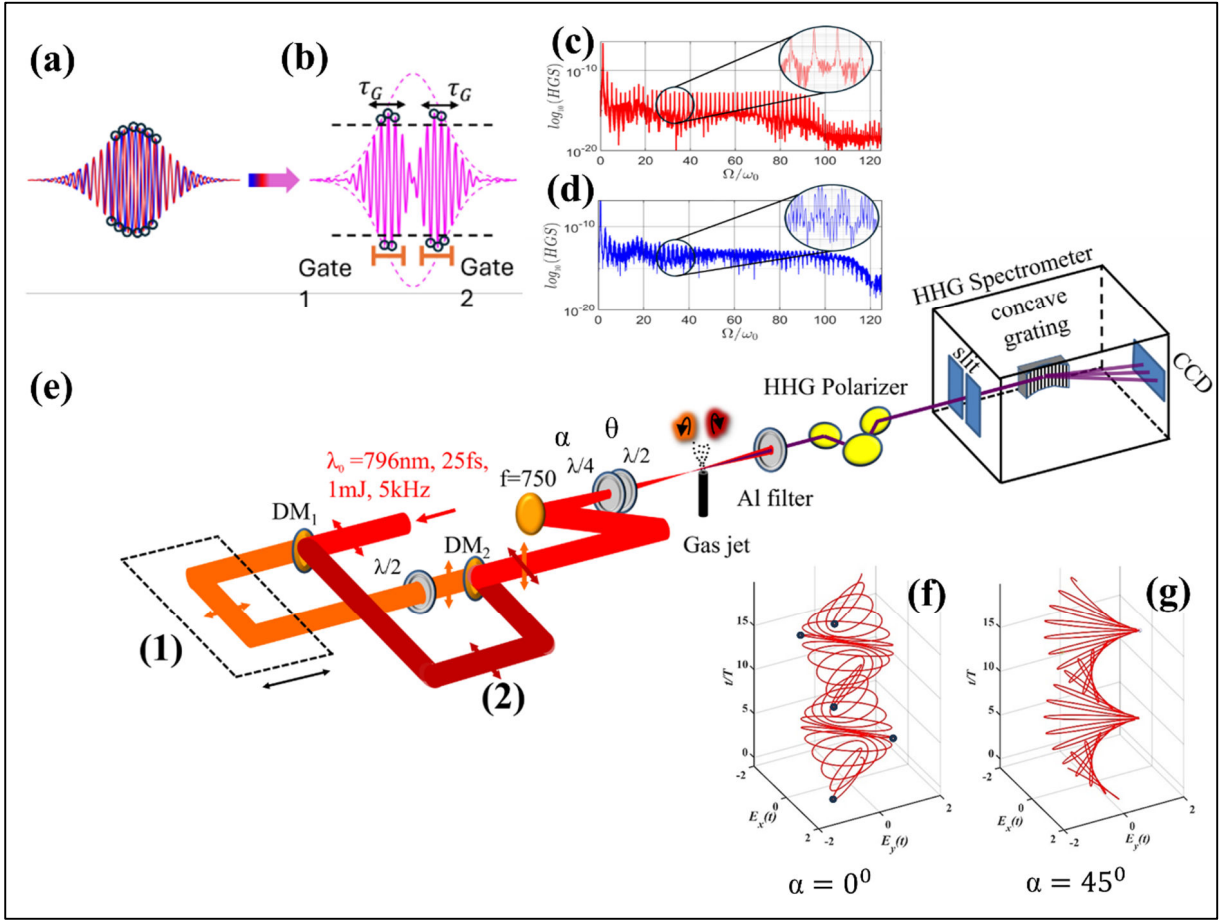

**Fig. 1: The concept behind the VTG scheme. (a)** in scalar time-gating (TG), the proximity of the central frequencies of a two-color co-linearly-polarized driver results in envelope modulation due to beating **(b)**, which effectively splits the pulse into sub-pulses ("gates"). Here two gates are shown, each encompassing 5 recollisions (black circles). The reduction in the number of recollisions (as compared to the monochromatic case), results in the generation of closely-spaced harmonic peaks, as seen in the blue HHG spectra (HGS) trace of panel **(d)**, as compared to the HGS with a monochromatic driver which is composed of odd-integer multiples of the fundamental frequency [red HGS trace of panel **(c)**]. In **(b)** two gates are seen, each encompassing several recollisions. **(e)** Experimental setup showing the interferometer [arm (1) and arm (2)], HHG Gold reflective polarizer and XUV spectrometer. DM-dichroic mirror. The half-waveplate in arm 1 rotates the polarization from horizontal to vertical. **(f)** Lissajous curve of a continuous-wave cross-linear TG vectorial driver, which is highly-elliptical at all times except for a serie of gates (symbolized here by the black dots) at which its polarization is close to linear. Consecutive gates (here 5 gates are shown) are cross-polarized as well and at those instants recollisions occurs. **(g)** The two cross-polarized fields can be further tailored using a quarter-waveplate, whose reading α dictates whether the total vectorial field at the focus remains at cross-linear configuration (α=0⁰), or is made counter-rotating cross-elliptical, or counter-rotating circular (α=±45⁰, Foucault-driver). The application of vectorial TG, together with the control imposed by this waveplate, changes the timings, directions and strength of the few recollisions which do take place, resulting in broadband harmonic emission, composed of highly-helical harmonic peaks, with high values of ellipticities and alternating helicities.

We start by describing analytically the outcome of the VTG scheme. The two-color cross-linearly-polarized fields are (see methods)

$\mathbf{E_1}(t;\alpha) = E_1 f(t - t_1) \cos(\omega_1 t) \mathbf{e_y}$,
$\mathbf{E_2}(t;\alpha) = E_2 f(t - t_2) \cos(\omega_2 t + \phi) \mathbf{e_x}$

The two colors have central frequencies $\omega_1 = \omega_0(1 + \delta), \omega_2 = \omega_0(1 - \delta)$ where $\delta$ is the normalized frequency difference $\delta = (\omega_1 - \omega_2)/2\omega_0 = (\omega_1 - \omega_2)/(\omega_1 + \omega_2)$ and $\omega_0 = (\omega_1 + \omega_2)/2$ is their average. $f(t - t_0)$ is a pulse envelope, centered around $t_0$, and $E_1, E_2$, are amplitudes. Passing these two fields through a quarter-waveplate whose reading is $\alpha = 0^0$ [the fast axis is at an angle $\alpha$ with respect to the vertical axis $\mathbf{e_y}$], and taking equal amplitudes $E_1 = E_2 = E_0$ we get that the total continuous-wave (CW) field is

$$\mathbf{E}(t; \alpha = 0^0) = E_0\sqrt{2} \begin{Bmatrix} \left[\cos\left(\omega_0 t + \frac{\phi}{2} - \frac{\pi}{4}\right)\cos\left(\omega_0\delta t - \frac{\phi}{2} + \frac{\pi}{4}\right)\right]\mathbf{e_{x'}} \\ \left[-\sin\left(\omega_0 t + \frac{\phi}{2} - \frac{\pi}{4}\right)\sin\left(\omega_0\delta t - \frac{\phi}{2} + \frac{\pi}{4}\right)\right]\mathbf{e_{y'}} \end{Bmatrix} \quad (1)$$

Where the perpendicular direction are along a reference frame along the intersection between $\mathbf{e_x}, \mathbf{e_y}$, namely $\sqrt{2}\mathbf{e_{x'}} = (\mathbf{e_x} + \mathbf{e_y})$, $\sqrt{2}\mathbf{e_{y'}} = (-\mathbf{e_x} + \mathbf{e_y})$. This CW field consists of a carrier wave of the average frequency $\omega_0$ and a modulated envelope, which gives a serie of gates. The Lissajous curve of this field is shown in Fig. 1(f). It exhibits time-dependent ellipticity where the driver's polarization is highly elliptical at all times, except for instances (inside the gates) at which it is close to linear. Those gates point in the alternating directions $\mathbf{e_{x'}}, \mathbf{e_{y'}}$, respectively. The shapes of the Lissajous curves at two consecutive gates show slightly different shape and hence two extremal types of gates exist:

**Type I**: $\mathbf{E}\left[t_{I,e_{x'}} = \frac{T}{2}k\frac{1}{1+\delta} \ ; \ \phi_I = \frac{\pi}{2}\left(1 + \frac{4k\delta}{1+\delta}\right) \ ; \ \alpha = 0^0\right] = E_0\sqrt{2}\mathbf{e_{x'}}$ , where $k \notin \mathbb{Z}$

**Type II**: $\mathbf{E}\left[t_{II,e_{y'}} = t_{I,e_{x'}} - \frac{T}{4\delta} \pm \frac{T}{4}; \ \phi_{II} = \phi_I \ ; \ \alpha = 0^0\right] = E_0\sqrt{2}\mathbf{e_{y'}}$

Where $T = 2\pi/\omega_0$. In case that $1/\delta$ can be written as an even number, the Lissajous curve in type I gate shows a cusp, since the carrier and peak of the gate are in phase. This is reminiscent of a "cos" pulse in the usual short-pulse terminology used in attosecond science to describe zero carrier-envelope-phase (CEP). Type-II gate is a "sin" pulse, with the field passing through the origin for $t = t_{I,e_{x'}} - \frac{T}{4\delta}$, and reaching local maxima/minima $\pm \frac{T}{4}$ after that time. At the two instants $t_{I,e_{x'}}, t_{II,e_{y'}}$ the field is locally-linearly-polarized and its amplitude is $\sqrt{2}E_0$. Hence, this type of driver introduces polarization-gates (PG) [Corkum1994, Platonenko1999, Tcherbakoff2003, Zair2004, Shan2005, Sola2006, Zhang2007, Feng2009, Li2019] into the recollision dynamics, restricting recollisions to occur at the gates only. Each two consecutive polarization-gates are separated by $\Delta t_{gates} = \frac{T}{4\delta}$ and the locally-linear fields at the gates are perpendicularly-polarized (along $\mathbf{e_x}, \mathbf{e_y}$,). When the field $\mathbf{E}(t; \alpha = 0^0)$ is used to drive HHG, a train of linearly-polarized attosecond pulses are generated in each gate. Since every two consecutive gates are perpendicularly-polarized, they don't interfere in the spectrum. However, two gates of the same type (separated by $2\Delta t_{gates}$) do interfere in the spectrum, which translates into interferences with separation between every two harmonic emission channels being $\Delta\Omega = \frac{2\pi}{2\Delta t_{gate}} = 2\delta\omega_0$ , in accordance with frequency-domain selection rules analysis [Ragonis 2024]. While two consecutive perpendicularly-polarized gates don't interfere in the spectrum, they do affect the ellipticity of the frequency-domain XUV field.

If the relative phase between the two colors $\phi$ is varied, this causes the entire field to merely shift along the time axis. The relative timings and orientations of the gates do not change. Hence the sequence and properties of the recollisions remain the same (time shift manifests itself only as a spectral phase of the HHG field). However, the pulse envelope $f(t)$ will dictate how many gates are included. The minimal number of gates which still allows helical XUV radiation to be generated is two. The driver pulses must not be shorter than $\Delta t_{gate}$ in order for two gates to be included. In addition the phase $\phi$ must be chosen such the two gates (and not just a single strong gate with two satellite weaker gates) are included. In our experiment $\delta \approx 0.0235$, $\Delta t_{gates} \approx 28.4fs$ , and since the pulse durations are 45fs and 55fs, respectively (see methods), a VPG field containing between 2 and 3 gates was used.

Figure 2 shows the results of corresponding Time-dependent Schrodinger equations (TDSE) numerical simulations of a single electron in a model potential of Argon atom (see methods). The two-colors are cross-linearly-polarized ($\alpha = 0^0$) and the phase $\phi$ was chosen such that a trapezoid 26.7fs would contain either 2 or 3 gates. Panel (b) shows the Gabor transform of a 2-gate pulse where 2 groups of recollisions are clearly seen. Broad HGS is obtained around each odd-harmonic peak [two harmonic orders are shown here, around $\Omega = 23\omega_0$ (c) and $\Omega = 27\omega_0$ (d), blue curves], and while the radiation is low-helical at harmonic channels $\Omega_{(n_1,n_2)} = (n_1 + n_2)\omega_0 + (n_1 - n_2)\omega_0\delta$ [Ragonis2024], it is highly-helical between the channels (where the spectral content is as high). When 3 gates are included, the ellipticity of the radiation between the channels reduces. If more gates are included (see supplementary Fig. S1) the entire XUV emission becomes linearly-polarized, as dictated by selection rules within the Floquet limit [Lerner2024]. As we will show, one way to break the dynamical symmetry obtained at the Floquet (CW) limit is to perturb the cross-linear configuration. Even a slight symmetry-breaking, by taking $\alpha = 1.5^0$, increase the ellipticity of the XUV radiation even if many gates are taken see supplementary Fig. S1).

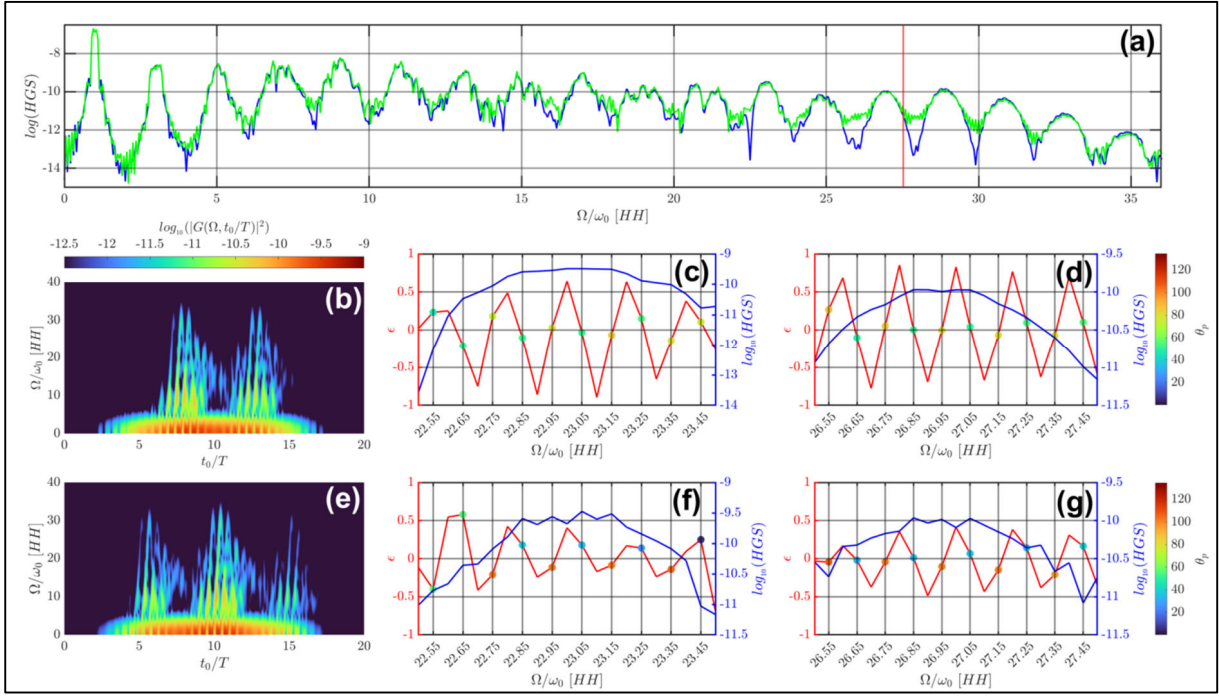

**Fig. 2:** Numerical TDSE results for the VTG scheme where the two colors are cross-linearly-polarized (α=0⁰) and δ=0.05. (a) HGS for a 26.7fs-long envelope containing 2 gates (blue curve) or 3 gates (green curve). (b)- Gabor transform for the 2-gate pulse (where the two gates, around t=7.5T and t=12.5T are clearly visible). (c) HGS intensity around $\Omega = 23\omega_0$ in log scale, log₁₀S(Ω) (right vertical axis, in blue) vs. harmonic order, and ellipticityi-helicity product $(\epsilon \cdot h)(\Omega)$ (left vertical axis, in red. Circles indicate low values of $(\epsilon \cdot h)(\Omega_{(n_1,n_2)})$ at the harmonic channels $\Omega_{(n_1,n_2)} = (n_1 + n_2)\omega_0 + (n_1 - n_2)\omega_0\delta$, predicted by Floquet analysis [Ragonis2024]. Between the channels the radiation is highly-helical. (d)-same as (c), but for $\Omega = 27\omega_0$ . (e-g)- same as (b-d) but for VTG pulse supporting 3 gates. As before the radiation at harmonic channels is low-helical, but this time between the channels not as highly-helical as in (c-d). This is due to the convergence to the Floquet limit.

Next, we check these predictions in experiment [see methods and also Figure 1(e) which presents a schematic sketch of the experimental set-up]. A two-color field with close frequencies was prepared by directing 25 fs pulses at a central wavelength of 796 nm from a commercial Ti:Sa laser to a Mach–Zehnder interferometer whose spectrally flat beam splitters were replaced with hard-edge short-pass dichroic mirrors with a cutoff wavelength at 795 nm [Ragonis2024] which resulted in two pulses of durations 45fs and 55fs, respectively (measured by SHG-FROG) and central wavelengths of $\lambda_1 =$ 777.8nm and $\lambda_2 =$ 815.2nm, respectively, yielding a detuning of $\delta \approx 0.0235$. At the exit of the interferometer the beams were spatially and temporally overlapped (arm 1 is placed on a translation stage) to yield maximal HHG signal, forming a two-color VTG driver. A super-achromatic half-wave plate (*B. Halle*) was placed in arm 1 of the interferometer, flipping the linear polarization from horizontal to vertical. With the value of detuning, $\Delta t_{gates} \approx 28.4 fs$, which (regarding the measured pulse durations) resulted in a VPG field containing between 2 and 3 gates.

In order to break the symmetry of the cross-polarized driver, the combined beam passed through a super-achromatic quarter-wave plate (*B. Halle*) placed on a motorized rotation stage prior to the HHG generation chamber. The orientation of this wave plate α controlled the polarization state and bandwidth [Ragonis2024] of the XUV radiation obtained, as the possible emission channels were dictated by the SAM conservation law. The two-color driver was focused (using a f = 750 mm focusing mirror), few millimeters before a 100μm-diameter nozzle Argon jet in order to favor the HHG emission from the short trajectories, yielding estimated intensities of $1.3 \cdot 10^{14}$ [W/cm$^2$] and $1.1 \cdot 10^{14}$ [W/cm$^2$]. The resulting XUV radiation was filtered from the remaining infrared driver radiation with a 0.2 μm-thick Aluminum foil. It then entered a home-built HHG XUV spectrometer containing a 20μm-entrance slit and having a typical resolution of 25 meV at 30 eV and 90 meV at 70 eV. Fig. 3a,b shows a typical α-scan, where the HGS was recorded as function of the quarter-waveplate reading α. The chessboard-like structure of the XUV emission could be explained by frequency-domain (Floquet) considerations, taking into account parity, energy and SAM conservation [Ragonis2024]. For cross-linear configuration ($\alpha=0^0$) a maximal number of allowed emission channels is obtained. These channels merge into a broad peak around each odd-harmonic. For $\alpha=\pm 45^0$ the Lissajous curve of the driver resembles a Foucault pendulum (Fig. 1g). Only 2 emission channels are allowed in this case, and the width of the emission peaks is minimal.

A polarization scan ($\theta$-scan) was performed by using a fixed, reflective polarizer composed of 3-bare gold mirrors and rotation of the two-color driver field in the polarization plane by a super-achromatic half-wave plate (*B. Halle*) placed after the quarter-waveplate, before the HHG generation chamber. By rotating the driver field in space, also the XUV radiation is rotated by the same amount. Fig. 3b shows the trace of transmitted XUV radiation, as function of the orientation of the half-waveplate $\theta$. By performing high-accuracy polarization scan, and fitting the transmitted trace to a Malus-type equation for each XUV energy, the ellipticity and polarization ellipse orientation angle which provided the best fit were retrieved. Of course, this type of measurement doesn't resolve the full Stokes parameters of the field, and is capable of providing an upper bound for the ellipticity $\epsilon(\Omega)$ only. The results of the experimental $\theta$-scan, for cross-linear configuration $\alpha=0^0$, are shown in Fig. 4 and are in accordance with the numerical results of Fig. 2. It is seen that at harmonic channels $\Omega_{(n_1,n_2)} = (n_1 + n_2)\omega_0 + (n_1 - n_2)\omega_0\delta$ predicted by Floquet analysis [Ragonis2024] the ellipticity of the XUV field low while between the channels the XUV field is highly-helical.

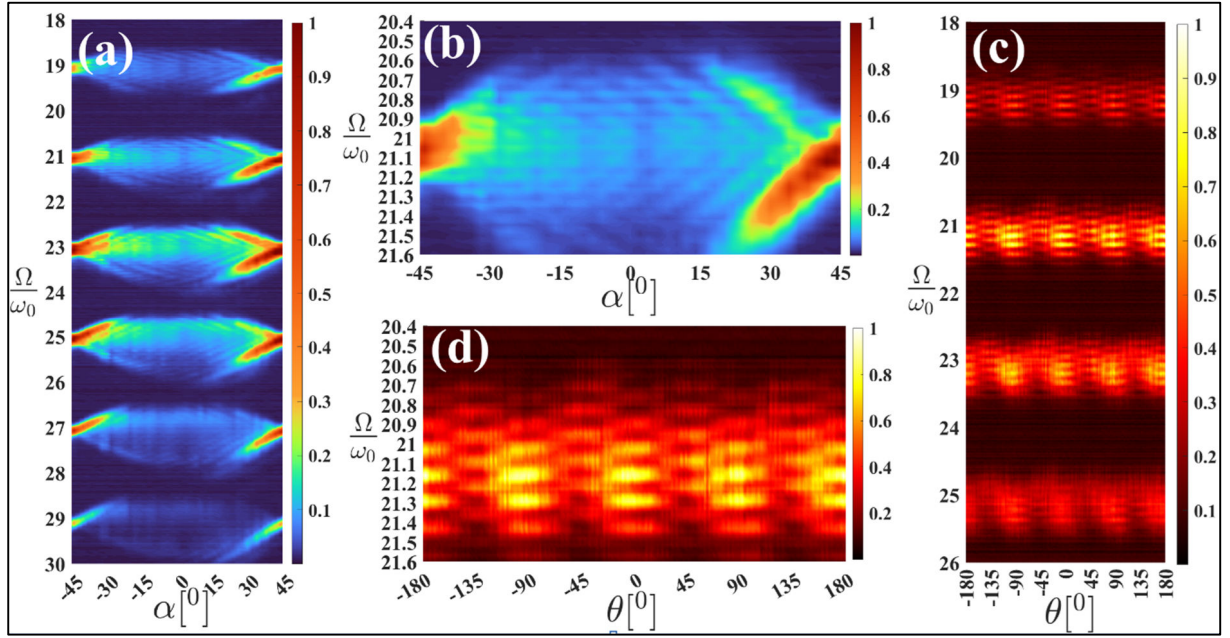

**Fig. 3:** Experimental results (raw data). (a) α-scan (b) zoom-in on the 21st harmonic. (c) θ-scan for α=0⁰. (d) zoom-in on the 21st harmonic. Measurements in panels (a,b) were taken with 20μm slit. Measurements in panels (c,d) were taken with 200μm slit and even so the alternations of the signal both as function of the half-waveplate reading $\theta$ and as function of the harmonic order are clearly seen.

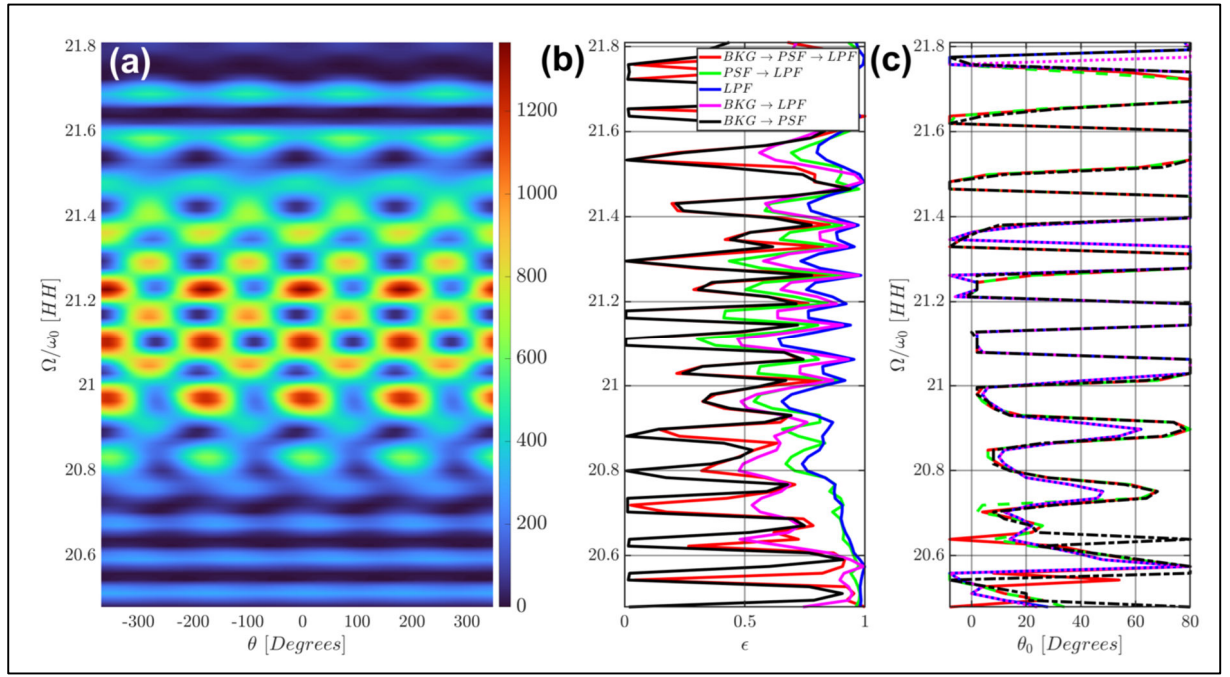

**Fig. 4:** Experimental polarization scan (θ-scan) for cross-linear configuration α=0⁰, 20μm slit, zoom-in on the 21st harmonic. In this scan the detuning was increased to δ=0.033 by shifting the central frequency of the laser.(a) raw data θ-scan after background substruction, deconvolution with the spectrometer's instrument response function (IRF), high-frequency noise removal by use of low-pass filtering (LPF) and fitting to a Malus-type formula. (b) the spectral ellipticity $\epsilon$ which results from the fitting algorithm shows periodic alternations between very low and very high values with period of 2δ between energies of low $\epsilon$, which are perpendicularly-polarized according to the values of their orientation ellipses $\theta_0$ (c).

In order to better understand the results of Fig. 4 we return to the numerical simulations and in Fig. 5 present full HGS α-scan maps $S(\Omega;\alpha)$ and ellipticity-scan maps $(\epsilon \cdot h)(\Omega;\alpha)$ for VTG driver supporting either 2 gates (panels a-d) or 3 gates (panels e-g), for $\Omega \approx 27\omega_0$. A clear periodic structure is seen both in the HGS and in the ellipticity-helicity maps. In the 2-gate case (panels a-d) the HGS map exhibits a fine-structure of 14 downward-tilted lines while in the 3-gate case (panels e-h) additional 14 upward-tilted lines are visible, forming a chessboard structure. This structure could be explained from a selection-rule perspective (i.e., assuming CW driver and a Floquet analysis) [Ragonis2024]. It is remarkable how the shift from 2 to 3 gates enhances the existence and applicability of the Floquet limit, despite of the fact that only about 6 recollisions (of the first and 3rd gates; see Fig. 2e) are interfering. Panels (c,g) show ellipticity-helicity maps $(\epsilon \cdot h)(\Omega;\alpha)$ which show the same tilted-lines/chessboard patterns. Panels (d,h) show zoomed-in view of panels (c,g). Within the Floquet limit it is predicted that in the cross-linear configuration (α=0⁰) the XUV radiation at harmonic channels $\Omega_{(n_1,n_2)} = 27\omega_0 + (n_1 - n_2)\omega_0\delta$ is linearly-polarized, along the same direction as the polarization of the driver color from which more photons were absorbed [Lerner2024]. Also here the XUV radiation at harmonic channels is close to being linearly-polarized while between the channels the ellipticity is higher. Breaking the cross-linear geometry, e.g., changing the value of α to α=1.5⁰ yields large ellipticity values at the harmonic channels.

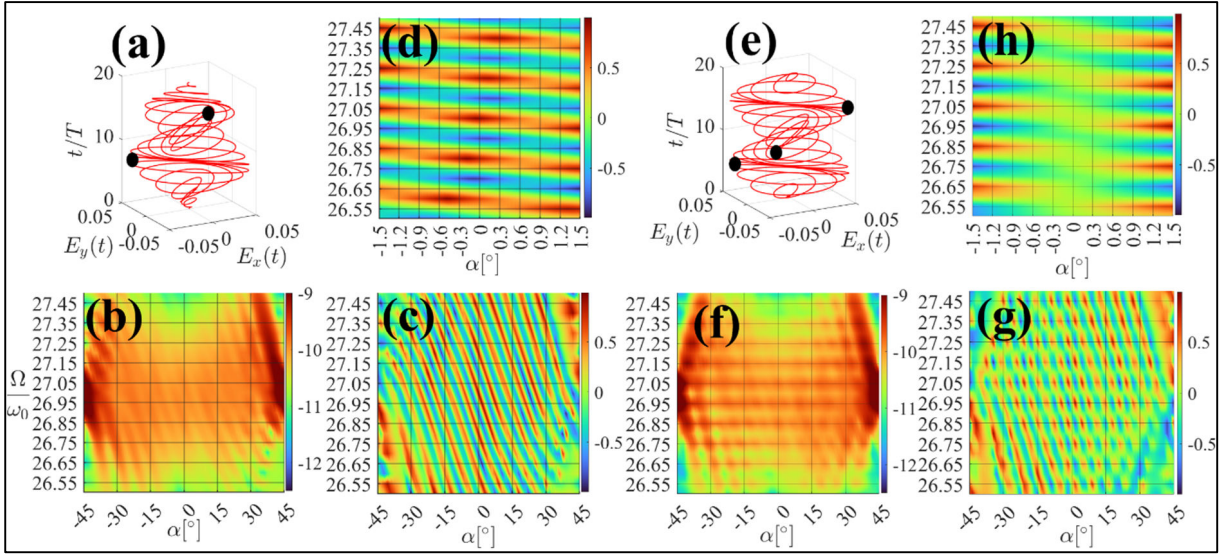

**Fig. 5**: Numerical simulations for VTG pulses supporting 2 and 3 gates (panels a and e, respectively, shown for cross-linear configuration α=0⁰), for the 27th harmonic. Panels (b,f) show HGS α-scans. Chessboard pattern, similar to the one obtained in the experiment, is obtained for 3 gates while for 2 gates only 14 downward-tilted lines are observed [Ragonis2024]. Panels (c,g) show ellipticity-helicity maps $(\epsilon \cdot h)(\Omega;\alpha)$ which show the tilted-lines/chessboard pattern as well. Panels (d,h) show zoomed-in view of panels (c,g). The horizontal black lines are at the energies of the harmonic channels $\Omega_{(n_1,n_2)} = 27\omega_0 + (2n-1)\omega_0\delta$, as predicted by Floquet analysis. Panels (d,h) clearly show that in the cross-linear configuration (α=0⁰) the XUV radiation at harmonic channels is close to linearly-polarized while between the channels the values of the ellipticities depend on whether 2 gates or 3 gates are taken (in which case the ellipticities are very large or just large). Regardless of the number of gates, breaking the cross-linear geometry, e.g., changing the value of α to α=1.5⁰ yields large ellipticity values at the harmonic channels.

Despite of strictly applicable only in the limit of inifinite-duration driver pulses, it is instructive to compare our results to the predictions of the Floquet theorem. In the Floquest limit, obtained with CW drivers, each of the two cross-elliptically-polarized drivers exhibits the dynamical-symmetry (DS) operator $\widehat{D}_f = \widehat{T} \cdot \widehat{\sigma}_f$ where $\widehat{T}$ stands for time-reversal and $\sigma_f$ is space reflection along the fast axis of the quarter-waveplate [Neufeld2019]. As such, all resulting harmonics should be elliptically-polarized

(with no knowledge on the actual value of the ellipticity) with major/minor axis corresponding to the reflection axis. For the cross-linear, two-color case, with infinite-long driver, the obtained channels should be linearly-polarized [Lerner2024, Shafir2009, Bordo2020]. However, our driver pulses are not infinitely long. As a result the Floquet predictions need not apply. For pulses in the VTP scheme this is especially true as α approached zero, since then not only does TG restrict recollisions, but the perpendicular orientations of adjacent gates further restricts the spectral interferences between these groups of recollisions. Hence we expect the predictions based on DS to become less and less correct. Surprisingly, the numerical and experimental results do support the Floquet prediction, which might be either accidental or demonstrate the applicability of Floquet limit even in short pulses, as has been demonstrated before [Fleischer2005, Lucchini2022]. In any case, since short pulses are used, the spectral content between the discrete channels becomes significant and this content has high elliptical polarization, which is the basis behind the VTP scheme. As $\alpha$ increases from zero towards the Foucault-driver case, more and more recollisions take place, and the picture starts to converge to the Floquest limit. In that case (i.e., with the Foucault driver), the system exhibits the DS $\hat{\tau}_2 \cdot \hat{R}_{\frac{2}{1+\delta}}$, for which no selection rules were derived yet based on the analysis of DS.

Lastly, we discuss the advantages of our scheme for X-ray magnetic CD (XMCD) experiments and sensitive detection of organic chiral molecules. First, with our proposed HHG scheme, a spectral resolution of roughly 50meV was demonstrated, in principle restricted only by the resolution of the XUV spectrometer. This is about 4-times better than in the renowned counter-rotating bicircular (ω,2ω) scheme [Baykusheva2018, Baykusheva2019] in which the resolution was limited to the spacing between 2 adjacent harmonic orders [e.g., 0.65eV with a (1900nm,950nm) driver]. It is also about 6-times higher as compared to table-top photoelectron CD (PECD) setups (see e.g. Fig.6 in [Beaulieu2016] and Figs.2-3 in [BeaulieuFerre2016]). Hence, our scheme is perfectly suited for resolving closely-spaced vibronic excitations, which are known to have a strong effect on the CD signal, offering a specific fingerprint of chirality. Second, our scheme offers at least 6-times better sensitivity over previous methods, since a typical HGS α-scan contains much more information than previously obtained scans, which were sparse. As can be seen in Fig.3a, but also in Fig.3 of [Ragonis2024], where 18fs and 26fs driver pulses were used, the portion of the plot where the signal is substantial (SNR>0) is about 45% of the total plot area. For comparison, in the counter-rotating bicircular scheme, a meaningful signal fills only few percent of the total map area: the spectral width of a single harmonic divided by the spacing between 2 nearby harmonics is roughly 0.05, and spin selection rules dictate a substantial HHG signal is obtained at about 0.25 of the quarter-waveplate angular support only, where the driver is truly close to being circular-counter-rotating, (see e.g. Fig 3 in [Baykusheva2019]. Hence, our scheme is 6-times [$\sqrt{45/1.25}$] more sensitive than the state of the art. Third, we have shown that at each XUV energy around odd-harmonic order 2$n$-1, the α-scan yields a periodic modulation of the polarization state, with period 2π/(4n-4). By switching to the Fourier domain, this deterministic feature could be used as a known reference and improve SNR, in the same way the controlled modulations provided by the acousto-optic modulator in typical CD experiments in the VIS aid in extracting accurate CD signals by lock-in detection. As such, our method offers not only higher sensitivity but also specificity between different chiral molecules, since the Fourier transformation of the HGS maps would provide a very distinct and characteristic pattern for different chiral molecules.

**Conclusions**

We have demonstrated a simple method to generate broadband, helical XUV radiation through table-top HHG setup, which is based on triple-symmetry-breaking of the most common HHG configuration which uses a linearly-polarized monochromatic driver. This basic configuration yields a sparse spectrum of linearly-polarized odd-integer harmonics. This scheme could be regarded as a degenerate limit of a two-color scheme, where the two colors are equal. Here we proposed a straightforward scheme to lift this degeneracy: first, by turning the single-color scheme into a two-color one (where the two colors are close in frequencies) the sparse HGS broadens. Second, by turning the two-color scheme from scalar to vectorial (by making the two colors cross-linearly-polarized), the XUV radiation becomes helical. Lastly, by adding ellipticity to both colors (by making the two colors cross-elliptically-polarized), the helical XUV radiation could be tuned. Hence, all in all, we perturbed the scalar single-color scheme and turned it into a vectorial two-color one. As is often the case with dynamical-symmetry-breaking, this results in rich, new spectral features in the XUV emission. Indeed, this integration of the time-gating and polarization-gating techniques generates well-controlled bursts of recollisions, occuring along different directions in the polarization plane, which results in a table-top HHG source capable of delivering XUV radiation which is both broadband and helical. Our source exhibits predictable modulations of the radiation ellipticity and helicity, without compromising the brightness, which makes it especially relevant for the detection of ultrafast dynamics in chiral matter. This is because the periodic pattern of helicity modulation constitute a reference against which chiral measurements could be performed, in the same way that lock-in measurements are done using acousto-optic modulators in the Visible regime. This should increase the sensitivity of our XUV source in the detection of chiral matter as compared to other methods. Since the spectral polarization could be easily tuned by slight symmetry-breaking induced by a quarter-waveplate, precise matching of the polarization state to a given spectral transition is enabled. Additional tuning is possible by changing the central frequencies of the two colors, e.g., by spectrally broadening the input pulses [Ragonis 2024] or by tuning the cutoff position at which the input spectrum is split by replacing the interferometer with a pulse-shaper. We anticipate that our scheme will facilitate studies of chiroptical phenomena is molecules and solids, enabling the interrogation of fine spectral features, restricted in principle only by the resolution of the XUV spectrometer.

**References**
[Suzuki2014] M. Suzuki, Y. Inubushi, M. Yabashi. and T. Ishikawa, "Polarization control of an X-ray free electron laser with a diamond phase retarder", J. Synchrot. Radiat. **21**, 466–472 (2014).

[Lutman2016] A. A. Lutman et al. "Polarization control in an X-ray free-electron laser", Nat. Photon. **10**, 468–472 (2016).

[Zayko2021] S. Zayko, O. Kfir, M. Heigl, M. Lohmann , M. Sivis, M. Albrecht and C. Ropers, "Ultrafast high-harmonic nanoscopy of magnetization dynamics", Nat. Comm. **12** 6337 (2021).

[Fan2015] T. Fan, P. Grychtol, R. Knut, C. Hernández-García, D. D. Hickstein, D. Zusin, C. Gentry, F. J. Dollar, C. A. Mancuso, C. W. Hogle, O. Kfir, D. Legut, K. Carva, J. L. Ellis, K. M. Dorney, C. Chen, O. G. Shpyrko, E. E. Fullerton, O. Cohen, P. M. Oppeneer, D. B. Milošević, A. Becker, A. A. Jaron Becker, T. Popmintchev, M. M. Murnanea and H. C. Kapteyn, "Bright circularly polarized soft X-ray high harmonics for X-ray magnetic circular dichroism", PNAS **112** 14206 (2015).

[Kfir2017] O. Kfir, S. Zayko, C. Nolte, M. Sivis, M. Möller, B. Hebler, S. S. P. K. Arekapudi, D. Steil, S. Schäfer, M. Albrecht, O. Cohen, S. Mathias and C. Ropers, "Nanoscale magnetic imaging using circularly polarizedhighharmonic radiation", Sci. Adv. **3** eaao4641 (2017).

[Willems2020] F. Willems, C. v. K. Schmising, C. Strüber, D. Schick, D. W. Engel, J.K. Dewhurst, P. Elliott, S. Sharma and S. Eisebitt, "Optical inter-site spin transfer probed by energy and spin-resolved transient absorption", Nat. Comm. **11** 871 (2020).

[Heinrich2021] T. Heinrich, M. Taucer, O. Kfir, P. B. Corkum, A. Staudte, C. Ropers and M. Sivis, "Chiral high-harmonic generation and spectroscopy on solid surfaces using polarization-tailored strong fields", Nat. Comm. **12** 3723 (2021).

[Azoury2019] D. Azoury, O. Kneller, M. Krüger, B. D. Bruner, O. Cohen, Y. Mairesse and N. Dudovich, "Interferometric attosecond lock-in measurement of extreme-ultraviolet circular dichroism", Nat. Photon. **13** 138 (2019).

[Huang2018] P.-C. Huang, C. Hernández-García, J.-T. Huang, P.-Y. Huang, C.-H. Lu, L. Rego, D. D. Hickstein, J. L. Ellis, A. Jaron-Becker, A. Becker, S.-D. Yang, C. G. Durfee5, L. Plaja, H. C. Kapteyn, M. M. Murnane, A. H. Kung and M.-C. Chen, "Polarization control of isolated high-harmonic pulses", Nat. Photon. **12** 349 (2018).

[Bengs2021] U. Bengs and N. Zhavoronkov, "Elliptically polarized high-harmonic radiation for production of isolated attosecond pulses", Sci. Rep. **11** 9570 (2021).

[Fleischer2014] A. Fleischer, O. Kfir, T.Diskin, P. Sidorenko and O. Cohen, "Spin angular momentum and tunable polarization in high-harmonic generation", Nat. Photon. **8** 543 (2014).

[Kfir2015] O. Kfir, P. Grychtol, E. Turgut, R. Knut, D. Zusin, D. Popmintchev, T. Popmintchev, H. Nembach, J. M. Shaw, A. Fleischer, H. Kapteyn, M. Murnane and O. Cohen, "Generation of bright phase-matched circularlypolarized extreme ultraviolet high harmonics", Nat. Photon. **9** 99 (2015).

[Hickstein2015] D. D. Hickstein, F. J. Dollar, P. Grychtol, J. L. Ellis, R. Knut , C. Hernández-García, D. Zusin, C. Gentry, J. M. Shaw, T. Fan, K. M. Dorney, A. Becker, A. Jaroń-Becker, H. C. Kapteyn, M. M. Murnane and C. G. Durfee, " Non-collinear generation of angularly isolated circularly polarized high harmonics", Nat. Photon **9** 743 (2015).

[Han2023] M. Han, J.-B. Ji, K. Ueda and H. W. Wörner, "Attosecond metrology in circular polarization", Optica **10** 1044 (2023).

[Beaulieu2018] S. Beaulieu, A. Comby, D. Descamps, B. Fabre , G. A. Garcia, R. Géneaux, A. G. Harvey, F. Légaré, Z. Mašin, L. Nahon, A. F. Ordonez, S. Petit, B. Pons, Y. Mairesse, O. Smirnova and V. Blanchet, "Photoexcitation circular dichroism in chiral Molecules", Nat. Phys. **14** 484 (2018).

[Ferre2015] A. Ferré, C. Handschin, M. Dumergue, F. Burgy, A. Comby, D. Descamps, B. Fabre, G. A. Garcia, R. Géneaux, L. Merceron, E. Mével, L. Nahon, S. Petit, B. Pons, D. Staedter, S. Weber, T. Ruchon, V. Blanchet and Y. Mairesse, "A table-top ultrashort light source in the extreme ultraviolet for circular dichroism experiments", Nat. Photon. **19** 93 (2015).

[Cireasa2015] R. Cireasa, A. E. Boguslavskiy, B. Pons, M. C. H. Wong, D. Descamps, S. Petit, H. Ruf, N. Thiré, A. Ferré, J. Suarez, J. Higuet, B. E. Schmidt, A. F. Alharbi, F. Légaré, V. Blanchet, B. Fabre, S. Patchkovskii, O. Smirnova, Y. Mairesse, V.R. Bhardwaj, "Probing molecular chirality on a sub-femtosecond timescale", Nat. Phys. **11** 654 (2015).

[Baykusheva2018] D. Baykusheva and H. J. Wörner, "Chiral Discrimination through Bielliptical High-Harmonic Spectroscopy", Phys. Rev. X **8** 031060 (2018).

[Baykusheva2019] D. Baykusheva, D. Zindela, V. Svoboda, E. Bommeli, M. Ochsner, A. Tehlar and H. J. Wörner, "Real-time probing of chirality during achemical reaction", Proc. Nat. Acad. Sci. **116** 23923 (2019).

[Fleischer2005] A. Fleischer and N. Moiseyev, "Adiabatic theorem for non-Hermitian time-dependent open systems", Phys. Rev. A **72** 032103 (2005).

[Lucchini2022] M. Lucchini, F. Medeghini, Y. Wu, F. Vismarra, R. Borrego-Varillas, A. Crego, F. Frassetto, L. Poletto, S. A. Sato, H. Hübener, U. De Giovannini, Á. Rubio and M. Nisoli, "Controlling Floquet states on ultrashort time scales", Nat. Comm. **13** 7103 (2022).

[Garcia2016] C. Hernandez-Garcıa, C. G. Durfee, D. D. Hickstein, T. Popmintchev, A. Meier, M. M. Murnane, H. C. Kapteyn, I. J. Sola, A. Jaron-Becker, and A. Becker, "Schemes for generation of isolated attosecond pulses of pure circular polarization", Phys. Rev. A **93** 043855 (2016).

[Khazanov2022] E. A. Khazanov, "Post-compression of femtosecond laser pulses using self-phase modulation", Quant. Elec. **52**, 208 (2022).

[Fleischer2006] A. Fleischer and N. Moiseyev, "Attosecond laser pulse synthesis using bichromatic high-order harmonic generation", Phys. Rev. A **74** 053806 (2006).

[Merdji2007] H. Merdji, T. Auguste, W. Boutu, J.-P. Caumes, B. Carré, T. Pfeifer, A. Jullien, D. M. Neumark and S. R. Leone, "Isolated attosecond pulses using a detuned second-harmonic field", Opt. Lett. **32** 3134 (2007).

[Lerner2024] G. Lerner, M. Even-Tzur, O. Neufeld, A. Fleischer and O. Cohen, "Reflection parity and sace-time parity photonic conservation laws in parametric nonlinear optics", **Phys. Rev. Res.** 6 L042034 (2024).

[Bordo2020] E. Bordo, O. Kfir, S. Zayko, O. Neufeld, A. Fleischer, C. Ropers and O. Cohen, "Interlocked attosecond pulse trains in slightly bielliptical high harmonic generation", J. Phys. Photonics **2** 034005 (2020).

[Corkum1994] P. B. Corkum, N. H. Burnett and M. Y. Ivanov, "Subfemtosecond pulses", Opt. Lett. **19** 1870 (1994).

[Platonenko1999] V. T. Platonenko and V. V. Strelkov, "Single attosecond soft-x-ray pulse generated with a limited laser beam", JOSA B **16** 435 (1999).

[Tcherbakoff2003] O. Tcherbakoff, E. Mével, D. Descamps, J. Plumridge and E. Constant, "Time-gated high-order harmonic generation", Phys. Rev. A **68** 043804 (2003).

[Zair2004] A. Zaïr, O. Tcherbakoff, E. Mével, D. Descamps, E. Constant, R. Lopez-Matens, J. Mauritsson, P. Johnsson and A. L'Huillier, "Time-resolved measurements of high order harmonics confined by polarization gating", Appl. Phys. B. 78 869 (2004).

[Shan2005] B. Shan, S. Ghimire and Z. Chang, "Generation of the attosecond extreme ultraviolet supercontinuum by a polarization gating", J. Mod. Opt **52** 277 (2005).

[Sola2006] I. J. Sola, E. Mével, L. Elouga, E. Constant, V. Strelkov, L. Poletto, P. Villoresi, E. Benedetti, J.-P. Caumes, S. Stagira, C. Vozzi, G. Sansone and M. Nisoli, "Controlling attosecond electron dynamics by phase stabilized polarization gating", Nat. Phys. **2** 319 (2006).

[Zhang2007] Z. Chang, "Controlling attosecond pulse generation with a double optical gating", Phys. Rev. A **76** 051403(R) (2007).

[Feng2009] X. Feng, S. Gilbertson, H. Mashiko, H. Wang, S. D. Khan, M. Chini, Y. Wu, K. Zhao, and Z. Chang, "Generation of Isolated Attosecond Pulses with 20 to 28 Femtosecond Lasers", Phys. Rev. Lett. **103** 183901 (2009).

[Li2019] J. Li, A. Chew, S. Hu, J. White, X. Ren, S. Han, Y. Yin, Y. Wang, Y. Wu and Z. Chang, "Double optical gating for generating high flux isolated attosecond pulses in the soft X-ray regime", Opt. Exp. **27** 30280 (2019).

[Ragonis2024] E. Ragonis, E. Ben-Arosh, L. Merensky and A. Fleischer, "Controlling the bandwidth of high harmonic emission peaks with the spectral polarization of the driver", Optics Letters **49** 2741 (2024).

[Neufeld2019] O. Neufeld, D. Podolsky and O. Cohen, "Floquet group theory and its application to selection rules in harmonic generation", Nat. Comm. **10** 405 (2019).

[Shafir2009] D. Shafir, Y. Mairesse, D. M. Villeneuve, P. B. Corkum and N. Dudovich, "Atomic wavefunctions probed through strong-field light–matter interaction,", Nat. Phys. **5** 412 (2009).

[Beaulieu2016] S. Beaulieu, A. Comby, B. Fabre, D. Descamps, A. Ferre, G. Garcia, R. Geneaux, F. Legare, L. Nahon, S. Petit, T. Ruchon, B. Pons, V. Blanchet and Yann Mairesse, "Probing ultrafast dynamics of chiral molecules using time-resolved photoelectron circular dichroism", Faraday Discuss. **194** 325 (2016).

[BeaulieuFerre2016] S. Beaulieu, A. Ferré, R. Géneaux, R. Canonge, D. Descamps, B. Fabre, N. Fedorov, F. Légaré, S. Petit, T. Ruchon, "Universality of photoelectron circular dichroism in the photoionization of chiral molecules", New J. Phys. **18** 102002 (2016).

[Henke1993] B.L. Henke, E.M. Gullikson, and J.C. Davis. *X-ray interactions: photoabsorption, scattering, transmission, and reflection at E=50-30000 eV, Z=1-92*, Atomic Data and Nuclear Data Tables Vol. **54** (no.2), 181-342 (July 1993).

**Acknowledgments**
E.B.A. acknowledges financial support of the quantum science and technology scholarship by the Israeli council of higher education. E.B.A., E.R. and A.F. acknowledge financial support from the Tel-Aviv University center of light-matter interactions. A.F. acknowledges support of Israel science foundation grant 529/19. The authors thank K. Itzhak, H. Assor, M. Azoulay and E. Rosen from the Tel-Aviv university chemistry machine shop for technical assistance.

**Author contributions**
A.F. conceived the idea. E.B.A., E.R. and A.F. carried out the numerical analysis. All authors carried out the experiments, and contributed to the analysis, interpretation and writing of the manuscript. A.F. supervised the project.

**Additional information**
Supplementary information is available in the online version of the paper. Reprints and permissions information is available online at www.nature.com/reprints. Correspondence and requests for materials should be addressed to A.F.

**Competing financial interests**
The authors declare no competing financial interests.

**Methods**
**Experimental set-up and measurement procedures.** Figure 1(e) presents a schematic sketch of the experimental set-up. A commercial amplified Ti:sapphire laser system (Legend Elite Duo HE+USX-5kHz by *Coherent*) operated at 5kHz repetition rate was used to deliver linearly-polarized (horizontal polarization) 25 fs pulses at a central wavelength of 796 nm (energy of $\omega_0$=1.557eV). 1.5mJ of the laser input pulses were spectrally split and recombined in a Mach–Zehnder interferometer whose spectrally flat beam splitters were replaced with hard-edge short-pass dichroic mirrors with a cutoff wavelength at 795 nm, The interferometer generated two pulses out of each incoming input pulse with central wavelengths of $\lambda_1$ = 777.8nm and $\lambda_2$ = 815.2nm, respectively, yielding a detuning of $\delta \approx$ 0.0235. At the exit of the interferometer the beams were spatially and temporally overlapped (arm 1 is placed on a translation stage, and the time delay was set not at zero but where the maximal HHG signal was obtained), forming a two-color driver. A super-achromatic half-wave plate (*B. Halle*) was placed in arm 1 of the interferometer, flipping the linear polarization from horizontal to vertical. In order to break the symmetry of the cross-polarized driver, the combined beam passed through a super-achromatic quarter-wave plate (*B. Halle*) placed on a motorized rotation stage prior to the HHG generation chamber. The orientation of this wave plate α controlled the polarization state and bandwidth [Ragonis2024] of the XUV radiation obtained, as the possible emission channels were dictated by the SAM conservation law. Using Second Harmonic-Generation Frequency-Resolved Optical Gating (SHG-FROG), the pulse durations from each arm just prior to the HHG chamber entrance window were measured to be 45 and 55 fs, respectively. The energy per pulse in each arm was 440 and 480 μJ, respectively. The two-color driver was focused (using a f = 750 mm focusing mirror), few millimeters before a 100μm-diameter nozzle Argon jet in order to favor the HHG emission from the short trajectories, yielding estimated intensities [derived from the cutoff location of the HGS, around the 39th harmonic] of $1.3 \cdot 10^{14}$ [W/cm$^2$] and $1.1 \cdot 10^{14}$ [W/cm$^2$]. The resulting XUV radiation was filtered from the remaining infrared driver radiation with a 0.2 μm-thick Aluminum foil. It then entered a home-built HHG spectrometer comprised of a 20μm-by-6 mm vertical entrance slit (located 1365 mm after the focus), aberration-corrected flat-field concave grating with 300 lines/mm (model 30-006 by *Shimadzu*), and a back-illuminated charge-coupled device (CCD camera (model Newton DO940P-BN by *Andor*). The spectrometer had a typical resolution of 25 meV at 30 eV and 90 meV at 70 eV. The XUV spectrometer was calibrated according to the odd-integer comb of harmonic order when HHG was driven by pulses coming directly from the laser. Possible small inaccuracies in the distances between the entrance slit, grating and CCD chip were taken into account, as well as possible tilt of the CCD chip, by performing a multidimensional fitting algorithm. The resulting calibration from pixel to eV is given by

$$eV = 32.3 \cdot exp(-0.003446 \cdot pixel) + 71.98 \cdot exp(-0.00102 \cdot pixel) + 9.098$$

Deconvolution (performed in the pixel dimension) was carried out, with Instrument Response Function chosen as $G(pixel) = exp\left[-4ln2\left((pixel - pixel_0)/\delta pixel\right)^2\right]$, with width δpixel=5. This width is the maximal one which yields a deconvolved HGS signal which is completely positive. The XUV spectrometer images the entrance slit onto the CCD. Hence, were the XUV spectrometer perfectly aligned, and a perfect monochromatic XUV source was used, we would have obtained δpixel=1.48 (20μm slit divided by 13.5μm pixel size of the CCD chip). We attribute this discrepancy to a non-perfect grazing angle on the grating, which, according to the specification of the grating, should be accurate to within $0.1^0$.

The polarization scan was performed by using a fixed, 3-bare gold mirror reflective polarizer and rotating the two-color driver field in the polarization plane. This was done by an additional super-achromatic half-wave plate (*B. Halle*) placed after the quarter-waveplate, before the HHG generation chamber. By rotating the driver field in space, also the XUV radiation is rotated by the same amount. For each XUV energy the trace of transmitted XUV radiation, as function of the orientation of the half-waveplate $\theta$, was fitted to a Malus-type equation, and the ellipticity which provided the best fit was retrieved:

$$I_T = A \frac{1}{1+\epsilon^2}\left[(R_p - R_s)(1-\epsilon^2)cos^2(2\theta - 2\theta_0) + (R_p\epsilon^2 + R_s)\right]$$

Where $A$ is a constant and $R_p, R_s$ are the Fresnel reflection coefficients for p-polarization (horizontal) and s-polarization (vertical) calculated as products for the 3 Gold surface (AOI=$67.5^0$, $45^0$, $67.5^0$), as taken from [Henke1993]. The polarizer gives a typical discrimination at 40eV of $R_s(40eV) = 0.032364869, R_p(40eV) = 0.001526712, R_s/R_p = 21.199066$.

The orientation of the polarization ellipse was given by $2\theta_0$.

**Numerical time-dependent Schrodinger equation simulations.**
Our numerical simulations are based on solving the three-dimensional time-dependent Schrödinger equation (3D-TDSE) of a single active electron in an effective potential of Argon, in the length gauge (atomic units are used)

$$i\hbar \frac{\partial}{\partial t}\Psi(\mathbf{r},t) = \left[-\frac{1}{2}\nabla^2 + V_{Ar}(r) + [x\mathbf{e_x} + y\mathbf{e_y}] \cdot [\mathbf{E_1}(t) + \mathbf{E_2}(t)]\right]\Psi(\mathbf{r},t)$$

Here $\mathbf{r}=(x,y,z)$ is the coordinate vector, are unit vectors along the x- and y-directions, respectively, and $\mathbf{E_1},\mathbf{E_2}$ are the vectorial electric fields of the laser. The long-range potential used $V_{Ar}(r) = -(1 + 0.2719 \cdot exp(-0.25r))/\sqrt{0.09192 + r^2}$ adjusted to fit the two lowest bound states of Argon, at -0.57915a.u. and -0.15476 a.u., respectively [Fleischer2013]. The ground state has a spherical symmetry of an *s*-shell, mimicking an ensemble of randomly-oriented *p*-shell Argon atoms. The electron is subjected to the two-color field (written as a 2-by-1 column vector, with the first entry being the x-component and the 2nd entry being the y-component:

$$\mathbf{E_1}(t;\alpha) = E_1 f(t - t_0)\begin{bmatrix} -sin(\alpha)\cos(\alpha)[\cos(\omega_1 t) - \sin(\omega_1 t)] \\ \cos^2(\alpha)\cos(\omega_1 t) + \sin^2(\alpha)\sin(\omega_1 t) \end{bmatrix}$$

$$\mathbf{E_2}(t;\alpha) = E_2 f(t - t_0)\begin{bmatrix} \sin^2(\alpha)\cos(\omega_2 t + \phi) + \cos^2(\alpha)\sin(\omega_2 t + \phi) \\ -\sin(\alpha)\cos(\alpha)[\cos(\omega_2 t + \phi) - \sin(\omega_2 t + \phi)] \end{bmatrix}$$

This field is obtained by taking the field exiting the interferometer $\mathbf{E_1}(t;\alpha) = E_1 f(t - t_0)\cos(\omega_1 t)\mathbf{e_y}$, $\mathbf{E_2}(t;\alpha) = E_2 f(t - t_0)\cos(\omega_2 t + \phi)\mathbf{e_x}$ and passing it through the quarter waveplate. A trapezois pulse envelope $f(t - t_0)$ is taken, whose constant part supports 10 or 15 optical cycles $T = 2\pi/\omega_0$ where $\omega_0 = 0.056954195 a.u.$ corresponds to a wavelength of 800nm

The two-color field $\mathbf{E_1},\mathbf{E_2}$ has central frequencies $\omega_1 = \omega_0(1 + \delta), \omega_2 = \omega_0(1 - \delta)$ where $\delta$ is the normalized frequency difference $\delta = (\omega_1 - \omega_2)/2\omega_0 = (\omega_1 - \omega_2)/(\omega_1 + \omega_2)$ and $\omega_0 = 0.056954195 a.u.$ corresponds to a wavelength of 800nm. Usually $\delta = 0.05$ and trapezoid envelopes $f(t - t_0)$ were used, and $E_1 = E_2 = 0.045 a.u.$, yielding cutoff harmonic at order of 27.5.

The TDSE was integrated using a 6th order Forest– Ruth split-operator method in 3D Cartesian grid, with grid spacing and temporal step size yielding converged results. The initial ground state was obtained by imaginary time propagation. A complex boundary absorber was applied close to the end of the simulation boundaries to avoid unwanted reflections. The emitted harmonic field is calculated using the dipole acceleration expectation value $\mathbf{a}(t) \propto \langle\Psi(\mathbf{r},t)| - \nabla V_{Ar}(\mathbf{r})|\Psi(\mathbf{r},t)\rangle$. The two components of the dipole acceleration Fourier transform
$\bar{a}_j(\Omega) = \int_{-\infty}^{\infty} dt\, exp(-i\Omega t)\, \mathbf{a}(t) \cdot \mathbf{e_j} \equiv |\bar{a}_j(\Omega)| exp[i\phi_j(\Omega)]$ (j=x,y)
or the windowed-FT (Gabor transformed) accelerations $\bar{a}_{G,j}(\Omega) = \int_{-\infty}^{\infty} dt\, exp(-[(t - t_G)/\tau_G]^2)\, exp(-i\Omega t)\, \mathbf{a}(t) \cdot \mathbf{e_j}$ are used for calculating the HHG spectra (HGS) $S(\Omega) \equiv |\bar{a}_x(\Omega)|^2 + |\bar{a}_y(\Omega)|^2$, the Gabor-transformed HGS $S_G(\Omega) \equiv |\bar{a}_{G,x}(\Omega)|^2 + |\bar{a}_{G,y}(\Omega)|^2$ and the harmonic polarization (ellipticity):

$$\epsilon = \sqrt{\frac{1-\eta}{1+\eta}} \quad , \quad \eta \equiv \frac{1}{S(\Omega)}\sqrt{S^2(\Omega) + 4|\bar{a}_x(\Omega)|^2 \cdot |\bar{a}_y(\Omega)|^2\{\cos^2[\phi_y(\Omega) - \phi_x(\Omega)] - 1\}}$$

The orientation of the polarization ellipse is given by

$$\theta_0 = \frac{1}{2}tg^{-1}\left(\frac{2|\bar{a}_x(\Omega)| \cdot |\bar{a}_y(\Omega)|}{|\bar{a}_x(\Omega)|^2 - |\bar{a}_y(\Omega)|^2}\cos[\phi_y(\Omega) - \phi_x(\Omega)]\right).$$